# Optimal Academic Plan Derived from Articulation Agreements: A Preliminary Experiment on Human-Generated and (Hypothetical) Algorithm-Generated Academic Plans

David Van Nguyen[1], Shayan Doroudi[1, 2], and Daniel A. Epstein[1]

[1] School of Information and Computer Sciences, University of California, Irvine, USA

[2] School of Education, University of California, Irvine, USA

## Author Note

We have no conflicts of interest to disclose. The experiment was preregistered at AsPredicted (https://aspredicted.org/mj88x.pdf). This paper and our other related paper draw from the same dataset. However, each paper presents distinct results. This paper uses the dataset's experimental data on optimal academic plans. Our other paper uses the dataset's non-experimental data on ASSIST software feature suggestions (Nguyen et al., 2024).

**David Van Nguyen:** Conceptualization, Funding Acquisition, Methodology (Lead), Investigation, Formal Analysis, Writing - Original Draft, Writing - Review & Editing (Lead). **Shayan Doroudi**: Supervision (Equal), Methodology (Supporting), Writing - Review & Editing (Supporting). **Daniel A. Epstein**: Supervision (Equal), Methodology (Supporting), Writing - Review & Editing (Supporting).

Correspondence should be addressed to David Nguyen. Email: dvnguye5@uci.edu

## Abstract

Our preliminary experiment examined a potential pain point with ASSIST, California's database of articulation agreements. That pain point is cross-referencing multiple articulation agreements to manually develop an *optimal* academic plan. Optimal is defined as the minimal set of community college courses that satisfy all transfer requirements for the multiple universities a student is preparing to apply to. Accordingly, we designed a low-fidelity prototype that lists the minimal set of courses a *hypothetical* optimization algorithm would output based on selected articulation agreements. 24 students were tasked with creating an optimal academic plan using either ASSIST (which requires manual optimization) or the optimization prototype (which already provides the minimal set of classes). Prototype users had less optimality mistakes, were faster, and provided higher usability ratings compared to ASSIST users. Going forward, future research needs to move beyond our *proof of value* of a hypothetical optimization algorithm and towards actually implementing an algorithm.

## Introduction

If available, community college students refer to articulation agreements to take the correct community college courses to be eligible to transfer to a four-year college. Articulation agreements are "formal arrangements that establish course equivalencies and the transferability of academic credit in an effort to facilitate seamless transfer of students' credit across postsecondary institutions" (Crisp, 2021, p. 61). However, submitting only one transfer application can be risky because there is a possibility of admissions rejection (Neault & Piland, 2014). Thus, community college students tend to submit transfer applications to multiple universities (Jabbar et al., 2021).

In an ideal world, state higher education systems would have one uniform set of community college major-related courses that would satisfy all lower-division major requirements for all public universities in the state (Hodara et al., 2017). However, many states have non-uniform lower-division major requirements among their in-state public universities (Education Commission of the States, 2022; Hodara et al., 2017). Non-uniform transfer requirements may have negative impacts on community college academic planning, excess community college course credits, and time-to-transfer (Dunmire et al., 2011; Lewis et al., 2016; Reddy & Ryan, 2021).

Accordingly, our study examined one pain point users may have with Articulation System Stimulating Interinstitutional Student Transfer (ASSIST), which is "the official statewide database [of articulation agreements] that shows … [how California] community college [courses] may be used to satisfy elective, general education and major requirements" at universities within the University of California (UC) and California State University (CSU) system (ASSIST, n.d., para. 2). That pain point is cross-referencing multiple articulation agreement reports to manually develop an *optimal* academic plan. We define optimal as the minimal set of community college courses that satisfy all the transfer requirements for the multiple universities a student is preparing to apply to.

To their credit, California has implemented policy reforms to streamline and unify transfer requirements across UC and CSU campuses such as the California General Education Transfer Curriculum (Cal-GETC), Associate Degree for Transfer, and UC Transfer Pathways. Despite these policy reforms, non-uniform transfer requirements still persist (Lewis et al., 2016; Mejia et al., 2023; Reddy & Ryan, 2021).

To address that pain point, we designed a low-fidelity prototype (Sefelin et al., 2003) that lists the minimal set of community college courses that a *hypothetical* optimization algorithm would output based on the user's selected articulation agreements. To be clear, the minimal set of courses within the low-fidelity prototype was *not* actually generated by a computer algorithm but was prepopulated by the authors. We then conducted a preliminary experiment comparing optimal academic plan development using the optimization prototype versus the status quo method with ASSIST.

To be transparent, the purpose of our low-fidelity prototype and preliminary experiment was to serve as a *proof of value* before investing substantial resources into research and development of a potentially non-effective software solution. For context, "Unlike a POC [proof of concept], which focuses on technical feasibility, a POV [proof of value] seeks to show how the [potential software] solution can … bring significant benefits to the company or end-users" (Humanperf Software, 2023, para. 6).

## Theoretical Framework

Our study's theoretical framework is a synthesis of two existing theoretical frameworks. The first is Scott-Clayton's (2011) community college *structure hypothesis*:

> community college students will be more likely to persist and succeed in programs that are tightly and consciously structured, with relatively little room for individuals to deviate on a whim—or even unintentionally—from paths toward completion, and with limited bureaucratic obstacles for students to circumnavigate. (p. 1)

The second is Reeping and Knight's (2021) framing of *information asymmetries* within the context of community college. They specifically explored how two factors – *fragmentation* and *language* – contributed to students having less information about the community college transfer process compared to more powerful stakeholders:

> Fragmentation captures the degree to which … information is all in one place or is "fragmented" into pieces across several [web]pages. Language refers to the linguistic characteristics of the information presented in the webpages. (p. 319)

We adapt Reeping and Knight's theoretical framework by adding a third factor: *formatting* of information. Formatting refers to design, layout, columns, boxes, colors, etc.

### Lack of Structure with Non-Uniform Lower-Division Major Requirements

Community college students and academic advisors report it can be difficult to fulfill differing lower-division major requirements for multiple universities (Blash et al., 2012; Hodara et al., 2017; Holliday-Millard, 2021; Karandjeff & Schiorring, 2011; Lewis et al., 2016; Schudde et al., 2021). For example, in Hodara et al.'s (2017) study "a community college advisor estimated that advisors had to keep track of approximately 1,280 articulation agreements because there are 16 public universities [in their state] and about 80 programs at each institution" (p.

341). Hodara et al. (2017) argues that non-uniform major requirements necessitates more individualized student advising, which consequently increases the likelihood of academic advisors making mistakes.

## Language, Formatting, and Fragmentation in Articulation Agreements

Regarding language, “the average articulation agreement is written at the 16.4th-grade reading level,” which may not be readable by community college students (Taylor, 2017, p. 3). Furthermore, articulation agreements often “uses language to instill doubt about the transferability or applicability of student credits” and also use undefined jargon (Reeping & Knight, 2021, p. 332).

Regarding formatting, users state that articulation agreement formatting can be confusing to interpret (Holliday-Millard, 2021; Taggart et al., 2000). Furthermore, users state it can be difficult to cross-reference multiple articulation agreements when each agreement has a different format (Holliday-Millard, 2021; Taggart et al., 2000).

Regarding fragmentation, users have a difficult time locating articulation agreements in their state university system when agreements are decentralized across different university websites (Holliday-Millard, 2021; Katsinas et al., 2016). Whereas students were able to easily find articulation agreements that were located within one centralized website (Taggart et al., 2000). Another type of fragmentation is users needing to cross-reference multiple articulation agreements to develop an academic plan (as opposed to referencing a single document).

## Cross-Referencing Reports to Manually Develop an Optimal Academic Plan

To envision the challenges involved with creating an optimal academic plan, imagine that a Glendale Community College student is interested in applying to the history major at UC San Diego (UCSD) and CSU Fullerton (CSUF). Note that we use fictitious ASSIST reports (see

Figure 1) to illustrate the cross-referencing process succinctly without getting bogged down by real ASSIST reports, which can be confusing for first-time viewers. As a simplifying assumption, we do not consider course prerequisites, course scheduling, or a student's course preferences.

**Figure 1**

*Major Requirement Variability in Fictitious ASSIST Reports*

**Glendale Community College to**
**University of California, San Diego**
**History Major Articulation Agreement**

| University Major Requirements | Approved Community College Equivalents |
| --- | --- |
| Writing Course | ENG 200<br>--- Or ---<br>ENG 240 |
| American History Course | HIST 50<br>--- Or ---<br>HIST 70<br>--- Or ---<br>HIST 90 |

**Glendale Community College to**
**California State University, Fullerton**
**History Major Articulation Agreement**

| University Major Requirements | Approved Community College Equivalents |
| --- | --- |
| Writing Course | ENG 200 |
| American History Course | HIST 70<br>--- Or ---<br>HIST 90<br>--- Or ---<br>HIST 110 |

*Note.* These two ASSIST reports are fictitious and are used for illustrative purposes only.

UCSD requires one of these community college writing courses: ENG 200 or ENG 240. Whereas CSUF mandates ENG 200. Thus, ENG 200 is optimal.

Next, UCSD requires one of these history courses: HIST 50, HIST 70, or HIST 90. Whereas CSUF requires one of the following: HIST 70, HIST 90, or HIST 110. Thus, the optimal choice is to take *either* HIST 70 or HIST 90.

To recap, an optimal academic plan requires two courses. First is ENG 200. Second is either HIST 70 or HIST 90.

However, a student can make optimization mistakes when developing their academic plan, which can lead to taking four courses. The student could *first* take ENG 240 and HIST 110, which *partially* fulfills major requirements at UCSD and CSUF. Afterwards, the student realizes they have unfulfilled major requirements. The student then must *also* take ENG 200 for CSUF's unfulfilled major requirement and another history course for UCSD's unfulfilled major requirement.

While this cross-referencing process may seem straightforward, it becomes more difficult when students (a) pursue majors with complex requirements, (b) apply to more than two universities, and/or (c) optimize between major requirements and general education requirements.

## Methodology

We ran a between-subjects experiment comparing optimal academic plan development using either ASSIST or the optimization prototype. Before running the experiment, we ran pilot test sessions with three transfer students. We revised the experiment based on the pilot feedback.

We preregistered our experiment at AsPredicted (https://aspredicted.org/mj88x.pdf). We followed all aspects of our study preregistration except our sample and sample size. We initially planned to recruit California community college students. However, online recruitment through social media groups was not feasible. Instead, we opted for community college transfer students. Furthermore, we initially planned on a sample size of 18 but revised it to 24. Rest assured, statistical analysis was only conducted *after* all data collection was completed.

**Participants**

We posted the call for study participants on CSU and UC campuses' subreddit forums. The three eligibility requirements include (a) at least 18 years old, (b) current community college transfer student at a UC or CSU campus, and (c) majoring in any subject except psychology. Study participants received a $15 gift card.

24 transfer students participated in the study between November 2021 to December 2021. Given 5-point unipolar response options, study participants rated themselves as "very knowledgeable" of ASSIST ($M$ = 4.0, $SD$ = 0.7). Other participant demographics are listed in Table 1.

**Table 1**

*Transfer Student Participant Demographics*

| Demographic | *n* | % |
|---|---|---|
| Gender | | |
| Female | 13 | 54 |
| Male | 11 | 46 |
| Race and ethnicity | | |
| Asian | 10 | 42 |
| Hispanic or Latino | 5 | 21 |
| Middle Eastern or North African | 1 | 4 |
| Native Hawaiian or Other Pacific Islander | 2 | 8 |
| White | 10 | 42 |
| Age | | |
| 19-22 years old | 20 | 83 |
| 23-26 years old | 2 | 8 |
| 27-30 years old | 2 | 8 |
| Years enrolled in community college | | |
| 1-2 years | 15 | 62 |
| 3-4 years | 8 | 33 |
| 5-6 years | 1 | 4 |

## Materials

### *Academic Plan Worksheet*

The academic plan worksheet provided instructions, an area to type the courses for their optimal academic plan, and (if needed) a page to use as scratch paper. Among other things, the instructions explained the hypothetical scenario for the academic plan: they are an incoming community college student at Orange Coast College (OCC) and are planning to submit transfer applications to UC Berkeley (UCB) and UC Los Angeles (UCLA) as a psychology major. The simplifying assumptions for the academic plan include they (a) have not completed any major requirements yet, (b) do not need to fulfill general education requirements, and (c) do not need to consider course prerequisites, course scheduling, or personal interest among course options.

### *ASSIST Reports*

For study participants randomly assigned to use ASSIST, we provided ASSIST articulation agreement reports for OCC to UCB psychology major (ASSIST, 2021a) and OCC to UCLA psychology major (ASSIST, 2021b).

### *Academic Plan Optimization Prototype*

We designed a low-fidelity prototype that lists the minimal set of community college courses that a *hypothetical* optimization algorithm would output based on the user's selected articulation agreements. To be clear, the minimal set of courses within the low-fidelity prototype was *not* actually generated by a computer algorithm but was prepopulated by the authors.

To make it experimentally comparable to ASSIST reports, the prototype's minimal set of courses is presented in a report format, specifically a single combined report format (see Figure 2). The low-fidelity prototype was created in Microsoft Word. Mimicking the other experimental group, the prototype has the following input: community college was OCC and university majors were UCB psychology and UCLA psychology.

**Figure 2**

*Academic Plan Optimization Prototype (i.e., "Combined ASSIST Articulation Report")*

## Combined ASSIST Articulation Report

| USER INPUTS |
| --- |
| **Community College Selected:**<br>Orange Coast College<br><br>**University/Major Pairs Selected:**<br>• University of California, Berkeley – Psychology, B.A.<br>• University of California, Los Angeles – Psychology/B.A. |

| REPORT COURSE REQUIREMENTS | | |
| --- | --- | --- |
| **Row Instructions** | **Community College Course Option(s)** | **Course Satisfies Which Transfer Requirement(s)** |
| Complete the course in this row. | PSYC A100 - Introduction to Psychology | • UC Berkeley – Psychology Major<br>• UC Los Angeles – Psychology Major |
| Complete the course in this row. | ANTH A185 - Physical Anthropology | • UC Berkeley – Psychology Major |
| Complete the course in this row. | BIOL A225 - Human Physiology | • UC Berkeley – Psychology Major<br>• UC Los Angeles – Psychology Major |
| Complete the course in this row. | PHIL A220 - Introduction to Symbolic Logic | • UC Berkeley – Psychology Major<br>• UC Los Angeles – Psychology Major |
| Complete the course in this row. | MATH A182H - Honors Calculus 1 and 2 | • UC Berkeley – Psychology Major<br>• UC Los Angeles – Psychology Major |

**Figure 2 (Continued)**

| Complete ONE of the course options listed in this row. | ANTH A100 - Cultural Anthropology<br>--- Or ---<br>ANTH A100H - Honors Cultural Anthropology<br>--- Or ---<br>ANTH A190 - Introduction to Linguistics<br>--- Or ---<br>PSCI A180 - American Government<br>--- Or ---<br>PSCI A180H - American Government Honors<br>--- Or ---<br>PSCI A185 - Comparative Politics<br>--- Or ---<br>PSCI A188 - Introduction to Political Theory<br>--- Or ---<br>SOC A100 - Introduction to Sociology<br>--- Or ---<br>SOC A100H - Introduction to Sociology Honors | • UC Berkeley – Psychology Major |
|---|---|---|
| Complete ONE of the course options listed in this row. | CHEM A110 - Introduction to Chemistry<br>--- Or ---<br>CHEM A130 - Preparation for General Chemistry<br>--- Or ---<br>CHEM A180 - General Chemistry A<br>--- Or ---<br>PHYS A110 - Conceptual Physics<br>--- Or ---<br>PHYS A120 - Algebra-Based Physics: Mechanics<br>--- Or ---<br>PHYS A185 - Calculus-Based Physics: Mechanics<br>--- Or ---<br>PHYS A130 - University Physics 1 | • UC Los Angeles – Psychology Major |

END OF REPORT

*Note.* For easier communication, we referred to the academic plan optimization prototype as "Combined ASSIST" to our study participants.

***Usability Survey Scale***

The usability survey scale ($\alpha = 0.93$, 10 questions) measured usability of creating an optimal academic plan with either ASSIST or the prototype (see Table 2).

**Research Session Procedure**

The experiment took place in a private Zoom video conference room. We provided participants with study materials through Google Drive.

First, we walked participants through the academic plan worksheet. When participants finished reading instructions on the academic plan worksheet, we asked study participants to explain what the term "optimal" meant in the context of this study. In response, we stated that their interpretation was correct, or we fixed mistaken interpretations.

Then participants were randomly assigned to either use the prototype or ASSIST to create an optimal academic plan. Participants input their plan into the academic plan worksheet. To reiterate, the optimization prototype already lists the minimal set of community college courses (see Figure 2). So, filling out the academic plan worksheet was simple for prototype users. They just needed to follow the prototype's row instructions and type the courses from the prototype into the worksheet. In contrast, ASSIST users had to cross-reference multiple articulation agreement reports to *manually* develop an optimal academic plan.

After completing their academic plan, they were given a Qualtrics survey link to complete the usability scale questions.

**Table 2**

*Items in Usability Survey Scale*

| Scale Item | Response Options [a] |
| --- | --- |
| How much time was required to construct an **optimal** academic plan using the [REPORT_TYPE]? | None at all (5) – A great deal (1) |
| How much mental effort was required to construct an **optimal** academic plan using the [REPORT_TYPE]? | None at all (5) – A great deal (1) |
| How intuitive was it to construct an **optimal** academic plan using the [REPORT_TYPE]? | Not intuitive at all (1) – Extremely intuitive (5) |
| Are you confident that you correctly constructed an **optimal** academic plan using the [REPORT_TYPE]? | Not confident at all (1) – Extremely confident (5) |
| How many mistakes do you think you made in your **optimized** academic plan? Mistakes are defined as including unnecessary excess courses or failing to include necessary courses. | None at all (5) – A great deal (1) |
| How much background knowledge on course transferring policies was required to construct an **optimal** academic plan using the [REPORT_TYPE]? Background knowledge refers to information **not** included in the ASSIST report. | None at all (5) – A great deal (1) |
| How much help would you need from a counselor in order to construct an **optimal** academic plan using the [REPORT_TYPE]? | None at all (5) – A great deal (1) |
| How frustrating was it to use the [REPORT_TYPE] to construct an **optimal** academic plan? | Not frustrating at all (5) – Extremely frustrating (1) |
| How tempted were you to stop and give up on constructing an **optimal** academic plan using the [REPORT_TYPE]? | Not tempted at all (5) – Extremely tempted (1) |
| How satisfied were you with using the [REPORT_TYPE] to construct an **optimal** academic plan? | Not satisfied at all (1) – Extremely satisfied (5) |

*Note.* Depending on random assignment, "[REPORT_TYPE]" was either "separate ASSIST reports" or "combined ASSIST report." ("combined ASSIST report" refers to the academic plan optimization prototype.) [a] The column shows the text and numeric score of the first and last response option. Each scale item uses 5-point unipolar response options.

## Analysis

We preplanned to conduct Welch's two-tailed *t*-test with a significance level of 0.05. We used R for statistical tests. The experiment has three dependent variables: speed, usability, and number of optimality mistakes. To elaborate, mistakes are defined as excluding necessary courses or including unnecessary excess courses.

## Limitations

Our paper has several noteworthy limitations. First, *transfer students* in our study (see Table 1) are not representative of the *community college student body*. Second, the prototype report’s styling is similar but not an exact replica of ASSIST reports’ styling (e.g., colors, font, size), which may introduce experimental confounds.

# Results

Table 3 provides full statistical results from the Welch’s two-tailed *t*-tests. In summary, optimization prototype users performed better than ASSIST users in terms of mistakes, speed, and usability. All differences were statistically significant ($p < 0.05$) and had large effect sizes ($d > 0.8$).

Note that prototype users made *zero* optimality mistakes. This is not surprising since the prototype already provides the minimal set of courses for participants to type into their academic plan worksheet.

**Table 3**

*Welch's t-test Results Comparing ASSIST to the Academic Plan Optimization Prototype*

| Variable | ASSIST | | Prototype | | *t* | *df* | *p* | *d* |
|---|---|---|---|---|---|---|---|---|
| | *M* | *SD* | *M* | *SD* | | | | |
| Mistakes | 3.33 | 5.02 | 0.00 | 0.00 | 2.30 | 11.00 | 0.042 | 0.94 |
| Speed | 11.29 | 4.49 | 3.89 | 1.50 | 5.41 | 13.42 | $p < 0.001$ | 2.21 |
| Usability | 3.28 | 0.76 | 4.20 | 0.80 | -2.88 | 21.95 | 0.009 | 1.17 |

*Note.* N = 24 (n = 12 for each experimental condition). The unit of measurements are as follows. Optimality mistakes are the number of excluded necessary courses and included unnecessary excess courses. Speed is minutes. Usability is the score of a multi-item survey scale.

## Discussion

Aligning with our theoretical framework, our preliminary experiment found that prototype users had less optimality mistakes in their academic plan, were faster in creating their plan, and provided higher usability ratings compared to ASSIST users. This outcome was expected because ASSIST reports *arguably* do not excel within our theoretical framework criteria: structure, language, formatting, and fragmentation. However, to be clear, our experiment cannot disentangle what specific theoretical framework criteria (if any) contributed to ASSIST's worse experiment outcomes. Our experiment can only determine causal relationships between the independent variable (ASSIST or the prototype) and dependent variables (mistakes, speed, and usability).

Our preliminary experiment results provide several practical implications. First, students can make mistakes in manually optimizing an academic plan. Second, in states with non-uniform transfer requirements, optimization software can potentially help students transfer with fewer unnecessary excess community college credits, which may consequently help students transfer in fewer semesters. Third, optimization software might save users' time from having to *manually* develop an optimal academic plan. Freeing up academic advisors' time is crucial given high student-to-advisor ratios within community colleges (Carlstrom & Miller, 2013). Academic advisors could then use their limited time on advising tasks that cannot be effectively automated. However, future research needs to move beyond our *proof of value* of a hypothetical optimization algorithm and towards actually implementing an algorithm.

But before developing academic plan optimization software, researchers and ASSIST administrators should consider conducting needs assessments, feasibility studies, and co-design sessions with stakeholders. For example, are academic advisors able to *manually* develop optimal academic plans – without any optimality mistakes – within the time limits of a single counseling appointment? Furthermore, stakeholders might disagree with our prototype's definition of optimality (i.e., minimal set of courses). Instead, stakeholders might want to optimize an academic plan using other goals like maximizing GPA (Xu et al., 2016) or balancing course difficulty (Lefranc & Joyner, 2020).

While academic plan optimization software might be ideal, securing bureaucratic approval and developing such software can potentially take years. As a partial solution in the meantime, community colleges and ASSIST should explicitly teach students how to manually develop optimal academic plans using ASSIST reports.

To be clear, we are *not* advocating that all students should only pursue an optimal academic plan. However, students should be given the knowledge to make informed decisions about enrolling in unnecessary excess courses.